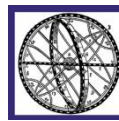

# Bronze Age sundial from Prokletije (Montenegro)

**Milos B. Petricevic [1], Larisa N. Vodolazhskaya[2]**

[1] Center for Conservation and Archaeology of Montenegro, Dept. of Archaeology, Kotor, Montenegro; E-mail: milos.b.petricevic@gmail.com

[2] Southern Federal University (SFU), Rostov-on-Don, Russian Federation; E-mails: larisavodol@aaatec.org, larisavodol@gmail.com

**Abstract**

The article presents the results of a study of signs on a Bronze Age slab "Sun stone" discovered at the foot of Maja e Can in Volušnica massif (Prokletije National Park, Montenegro). Studies have shown that the slab is an analemmatic sundial. The "Sun Stone" is more similar in marking to the slabs of the Srubnaya culture: all the cup marks are small and located along the ellipse line. Ellipses of bowl-shaped signs of analemmatic sundials from the Northern Black Sea region are similar in size to the reconstructed ellipse of the "Sun Stone". In addition, on one sundial from the Northern Black Sea region, the groove marks the distance that the gnomon must travel on the day of the winter solstice, similar to the groove on the Sun Stone. In the hour marking of sundial slabs from the Northern Black Sea region and Western Balkans, continuity can be traced, confirming the dating of the XV-XII centuries BC and indicating contacts of representatives of the synchronous local Glasinac culture (Glasinac II – Glasinac III) of the Western Balkans with representatives of the Srubnaya culture of the Northern Black Sea region.

**Keywords:** cup marks, hour markers, grooves, slab, analemmatic sundial, hour line, Bronze Age, Sun Stone.

In 2021, an employee of the Center for Conservation and Archaeology of Montenegro (Kotor) Miloš B. Petričević found out sandstone slab with petroglyphs at the foot of Maja e Can in Volušnica massif (Prokletije National Park) at 1710 m a.s.l. He asked the locals about petroglyphs and one of them, Ilijaz Duraković (Gusinje), said that he saw a stone in the mountains with the image of the Sun on it. He called it "Sun Stone". Ilijaz Duraković showed the location of this stone and Miloš B. Petričević was able to take photographs and explore it (Fig.1).

The stone slab with time markings is located in the open, in close vicinity to the nearby stream at the foot of Maja e Can in Prokletije National Park. The slab is situated approximately 500 m from well-known engraved rock site called " Šareni kamen". This is one of the few engraved rocks in the area of Volušnica massif (Fig. 2).

It shows a large number of anthropomorphic schematic figures dated back to Middle Bronze Age (Mijović, 1992, p. 23). The longest preserved part of the rock is 219 cm and the widest 170 cm, while the highest part of the stone block measured from the ground to the top is 56 cm (Fig. 3).

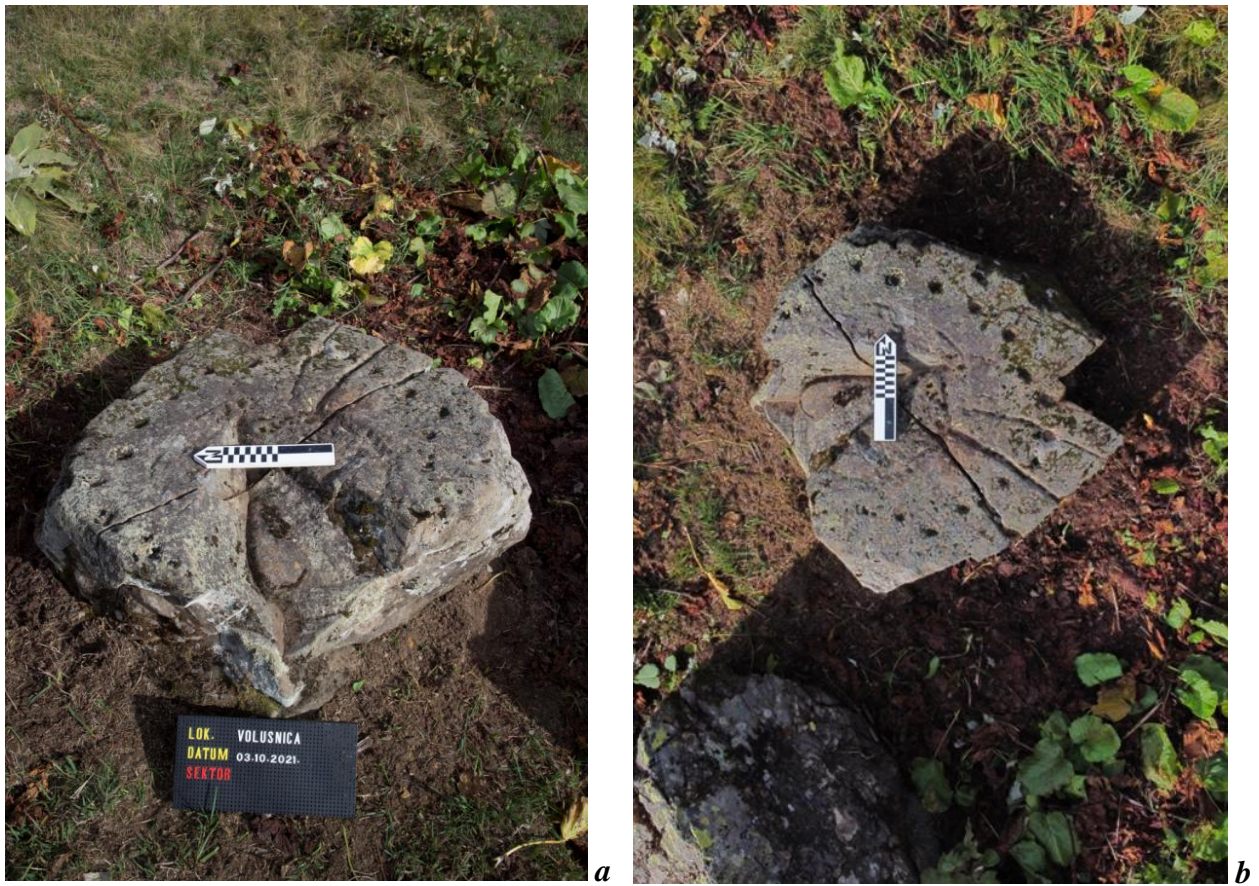

**Figure 1.** Montenegro, Prokletije National Park, foot of Maja e Can, photo of sandstone slab "Sun Stone" with petroglyphs at the discovery site: *a* – side view; *b* – top view. Photo by Miloš B. Petričević, 2021.

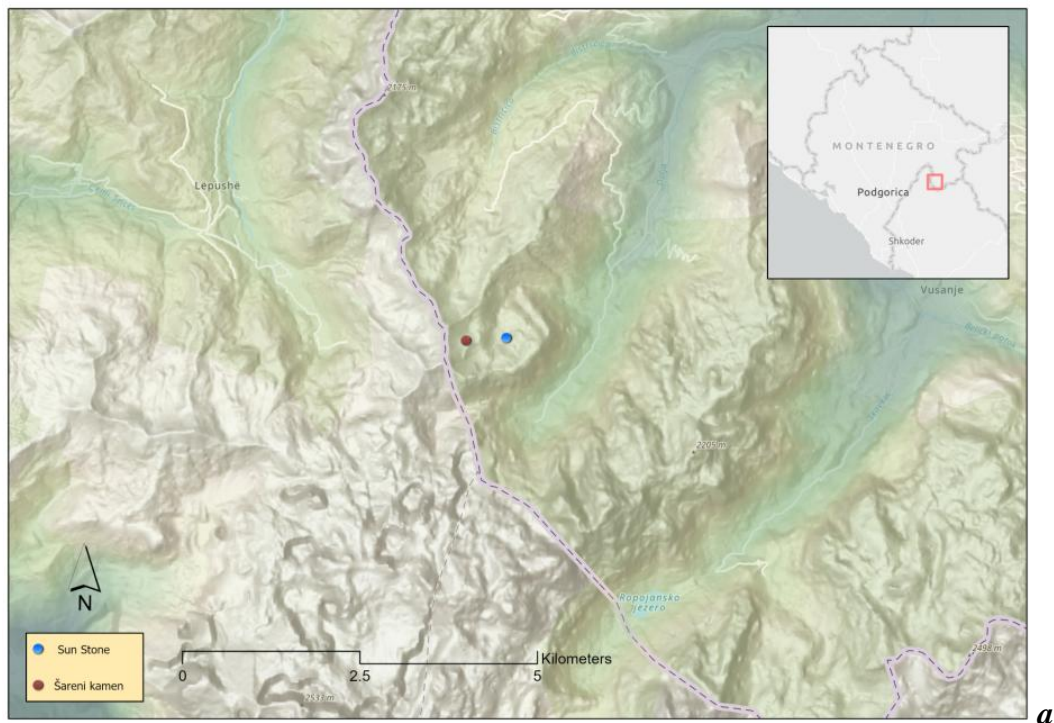

**Figure 2.** Map of Volušnica massif with marked places of discovery "Sun stone" (blue mark) and "Šareni kamen" (red mark).

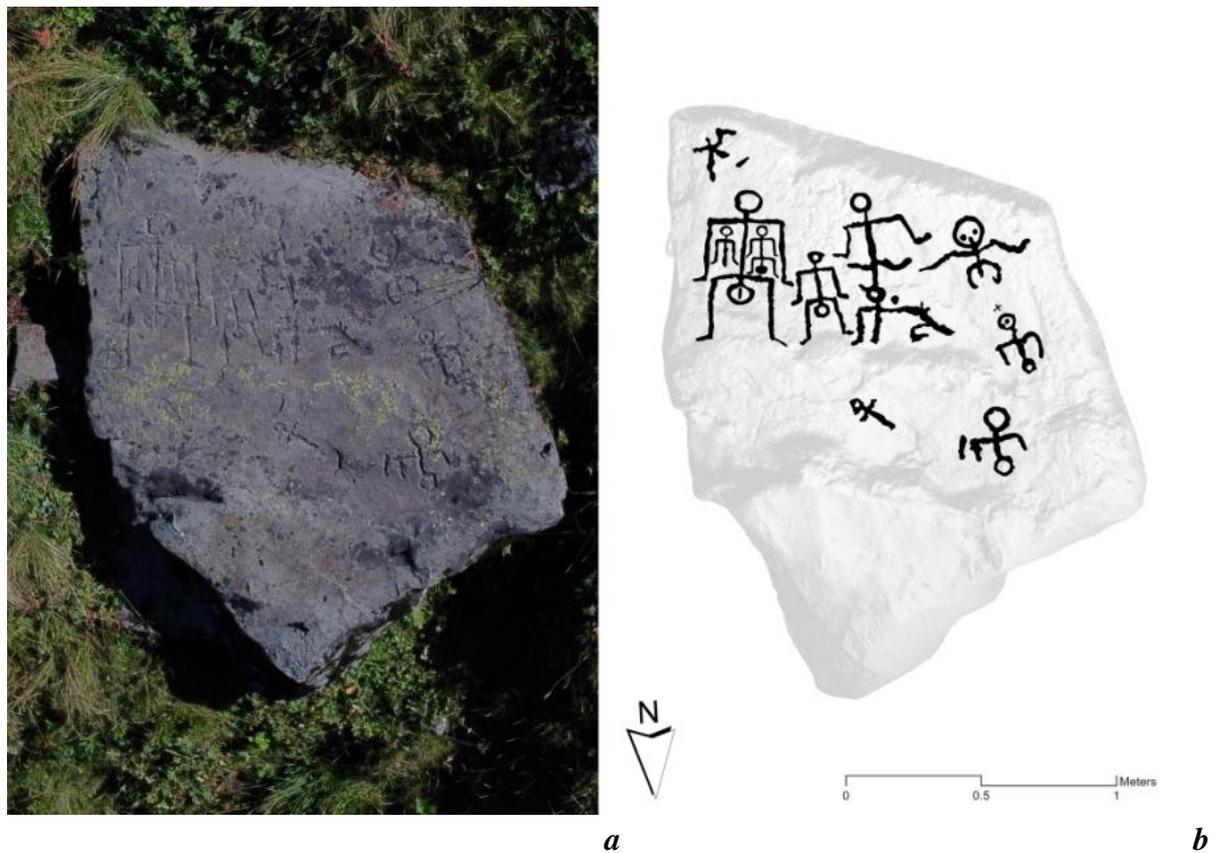

**Figure 3.** Slab with petroglyphs in the site "Shareni kamen": *a* – photo of the slab from above (Photo by Miloš B. Petričević, 2021); *b* – drawing petroglyphs on a slab (Miloš B. Petričević, 2021).

Prokletije hasal ways represented a serious scientific challenge to many research topics.

The area of Montenegrin high mountains and terrain, also known as Montenegrin Hills, mainly encompasses the northern part of Montenegro. This is an area characterized by vast limestone formations and plateaus bordered with several parallel ranges. This is an area of the highest mountains in the Dinaric Alps.

Heightwise, Prokletije is the biggest mountain chain of the southeastern Dinaric Alps which is not easy to spatially confine. The existing relief was, beside tectonic synthesis, greatly affected by exogenic forces, primarily glacial, fluvio-glacial, lime and fluvial erosion and denudation (Dragović, 1994, p. 77). The traces of Pleistocene glaciation in Prokletije prove that this area was one of the largest glacial zones on the Balkan peninsula.

Precisely because of its inaccessibility and spatial isolation, Prokletije still remains an uncharted territory for Montenegrin archaeologists. The British explorer sir Arthur Evans was the first archaeologist to visit Valbona Valley in search for archaeological remains (Evans, 1880, p. 10). The presence of archaeological remains in the region of Prokletije was first mentioned in the works of a famous Yugoslav anthropogeographer and ethnologist B. Gushich (Gušić, 1960, p. 388; 1964, p. 58). He was the first researcher who, at the beginning of the 20th century, explored the furthest this hardly accessible and hostile region. The most important archaeological discovery in the Montenegrin part of Prokletije are the petroglyphs located in the rock shelter called Vezirova brada (Vasić, 1972) and in Volušnica massif (Mijović, 1992).

Based on the condition and position of the rock it is impossible to establish whether the slab with petroglyphs at the foot of Maja e Can was found in its original place or position (in situ) concerning the azimuthal angle to geographic north. The longest and widest preserved parts of the sundial have approximately same dimensions, i.e. 94 cm and 93 cm respectively, while the highest part of the stone block measured from the ground to the top is 32 cm.

Long-term exposure to exogenic factors in a climate of snow, ice and extreme temperature fluctuations resulted in mechanical weathering and breakdown of the rock. This process, also known as frost wedging, represents the most common cause of rock weathering which occurs when the water filling the cracks in a rock freezes.

There is a crack on the surface of the the slab which caused the rock to physically break into two unequal parts while devastating the central petroglyph in the form of the Sun. There are indications that the horizontal surface containing the cup marks is not entirely preserved. This assumption is supported by the regular fractures on the edges which resulted from vertical cracks in the rock, i.e. decreased volume of the horizontal surface. On the surface of the slab there are visible sequenced cup-marks, a total of fifteen (15), which form an ellipse with a circular central motif in the form of an image of the sun and three radial grooves. The cup marks vary in diameter from 3.4 cm to 5.7 cm and depth from 0.9 cm to 2.3 cm.

For a more accurate photo fixation of the slab with petroglyphs, Miloš B. Petričević created a digital 3D model of the "Sun Stone". The Panasonic DMC-G5 digital camera was used to collect photographic images. To create a three-dimensional digital model of the slab with petroglyphs, the Agisoft Metashape program was used.

The engravings on the sundial's horizontal surface were analyzed by using computer programs (ArcGIS Pro and Global Mapper) in order to simulate various daylight conditions. The level and angle of light is a very important factor affecting the capability of the human eye to perceive contrast. Thus, the manipulation of the values of horizontal and vertical angles, i.e. angle height of the source of light increases the visibility of the engravings. The slope analysis based on the elevation differences was used in the process of identifying contour borders of the engravings.

A program called Relief Visualization Toolbox (RVT) specialized for visualization of digital surface models of ALS (Airborne Laser Scanning) data was used to increase visibility of the engravings (Kokalj, Zakšek, Oštir, 2011). In addition to this, the generated DSM (Digital Surface Model) model was used to collect information on the dimensions and cross section of the engravings (Fig. 4).

Traceology, as a form of functional analysis, studies the event traces created on lithic, bone and metal tools and can be, however, applied vice versa (Srejović 1997, p. 299). The identification of traces left by tools is nowadays at the higher level precisely due to techno-traceological studies in rock art (Zotkina, Kovalev, 2019). The motifs on the "Sun Stone" in Prokletije were created using pecking and abrasion techniques. The pecking method represents a technique of either direct or indirect removal of small portions of stone with lithic, bone or metal tools creating small crushed areas or "dints" (Keyser, Rabiega, 1999, p. 125). The abrasion technique implies a continuous direct contact between the tool (abrader) and surface, i.e. mechanical wear of the surface due to a directed frictional force. The nature of these traces is dependent on the physical features of the tool, more specifically its shape, size, weight and endurance.

Radial engravings on the slab were made using the abrasion technique. Three radial grooves are visible on the surface of the slab. It is not certain whether there was really first groove (bottom in figure 4) or the orientation of the crack in the stone block was actually caused by the direction of the engraving. The middle groove is slightly curved (30 cm long, 2 cm wide and 1 cm deep). The third groove (upper in figure 4) is noticeably shallower and has a more prominent curved form (35 cm long, 1.6 cm wide, 0.5 cm deep). A circular motif in the form of the Sun with a diameter of 44 cm, mostly destroyed by the weathering of rocks, is located in the center of the slab and the ellipse of holes.

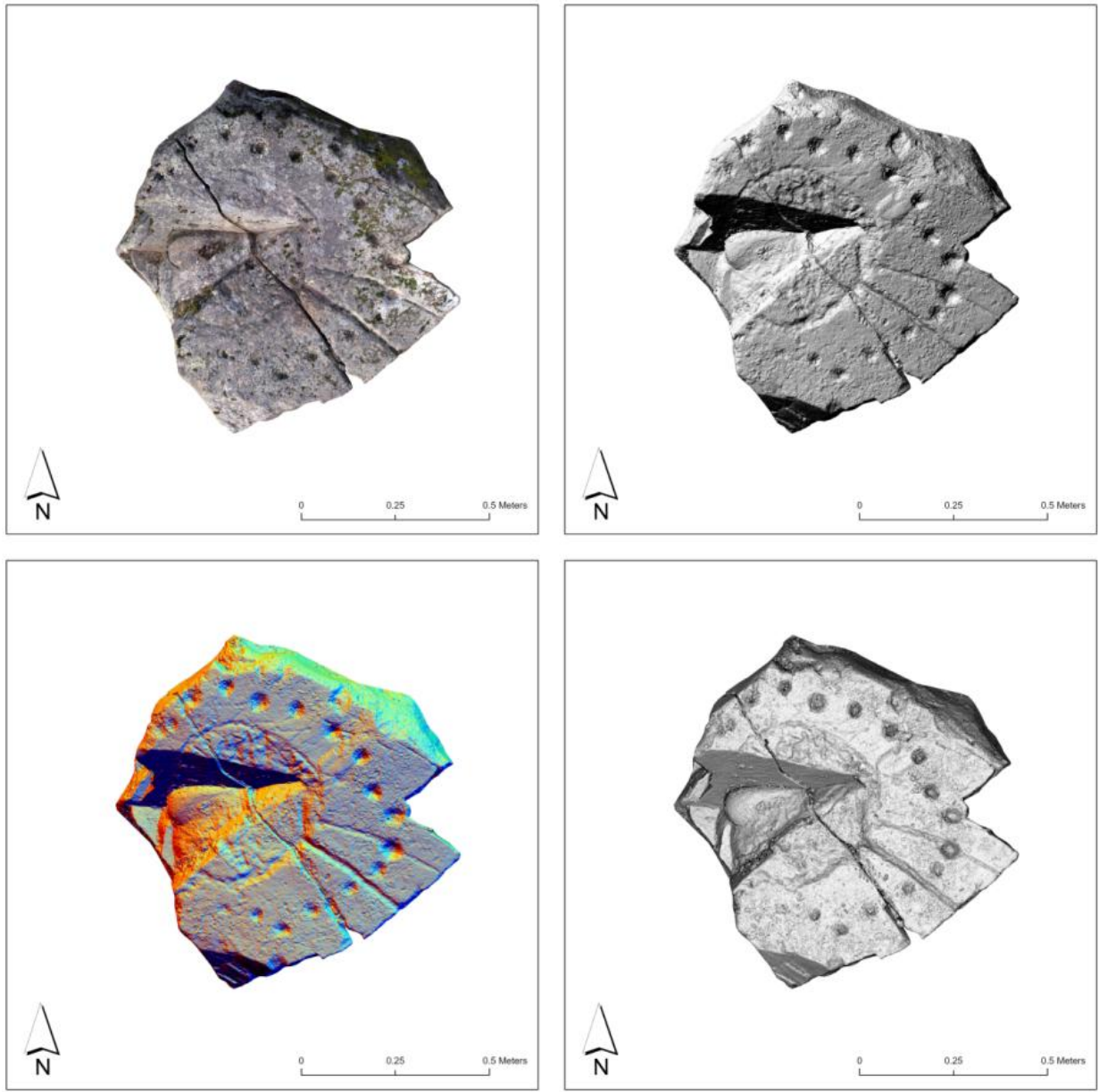

**Figure 4.** Orthophoto (upper left corner) and visualization of elevation data using Relief Visualization Toolbox – RVT.

The central circular motif and cup-marks were made using the pecking technique but for now we cannot claim with certainty whether it was a direct or indirect percussion. Peck traces are poorly distinguishable with the naked eye due visible bioactivity on the surface of the slab in the form of lichen and moss expansion, which prevents a more precise examination. Preliminary traceological analysis indicates that the interventions on the stone were done by employing different techniques and tools. We will be able to provide more specific information after carrying out a macroscopic and microscopic analysis of the pecking traces and engravings.

The main feature of the "Sun Stone" discovered in Prokletije are small cup marks located in a circle along the edge of the slab and a large drawn circle in the center of the slab, approximately from the center of which three grooves extend to the edge of the slab.

To date, three slabs are known with similar cup mark patterns. They were discovered in the Northern Black Sea region and date back to the Bronze Age. One slab attributed to the Dolmen culture, and the other two slabs attributed to the Srubnaya culture[1].

On the slab with petroglyphs, the holes were arranged in an ellipse, so Miloš B. Petričević suggested that the slab could be a sundial. He turned to L. Vodolazhskaya with a request to analyze the petroglyphs on the slab for a sundial.

A slab with cup marks from the Srubnaya culture burial of mound 1 of the Tavria-1 burial ground in the Rostov region is kept in the Archaeological Museum-Reserve "Tanais" (Rostov Region, Russia) (Larenok, 1998, p. 62) (Fig. 5). It dates back to the XV-XIII centuries BC and attributed to the Srubnaya culture.

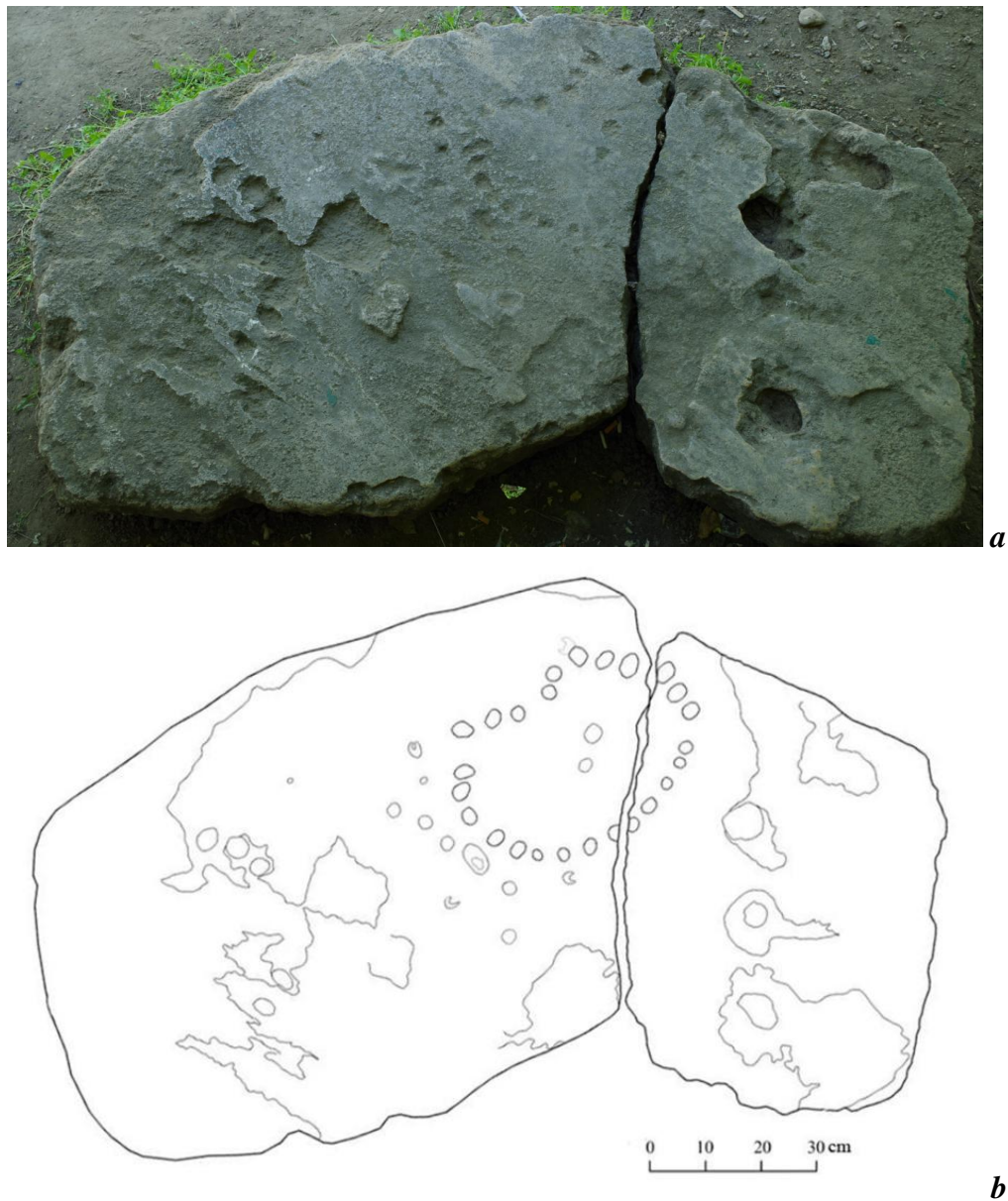

**Figure 5.** Slab with cup marks from the Tavria-1 burial ground in Rostov region: *a* – photograph of the surface of the slab with cup marks (Vodolazhskaya, Larenok, Nevsky, 2014, fig. 3); *b* – drawing the surface of a slab with cup marks (Vodolazhskaya, Larenok, Nevsky, 2014, fig. 4).

---

[1] Sometimes used the name "Srubna" culture.

The second slab is from the Srubnaya culture burial of mound 3 of the Popov Yar-2 mound group in the Donetsk region The second slab – from the Srubnaya culture burial of mound 3 of the Popov Yar-2 mound group in the Donetsk region is kept in the Donetsk Region History Museum (Donetsk region, Ukraine) (Polidovych, Usachuk, 2013; 2015, p. 444, 455; Polidovich et al., 2013, p. 59-62) (Fig. 6). It dates back to the XIII-XII centuries BC and also attributed to the Srubnaya culture.

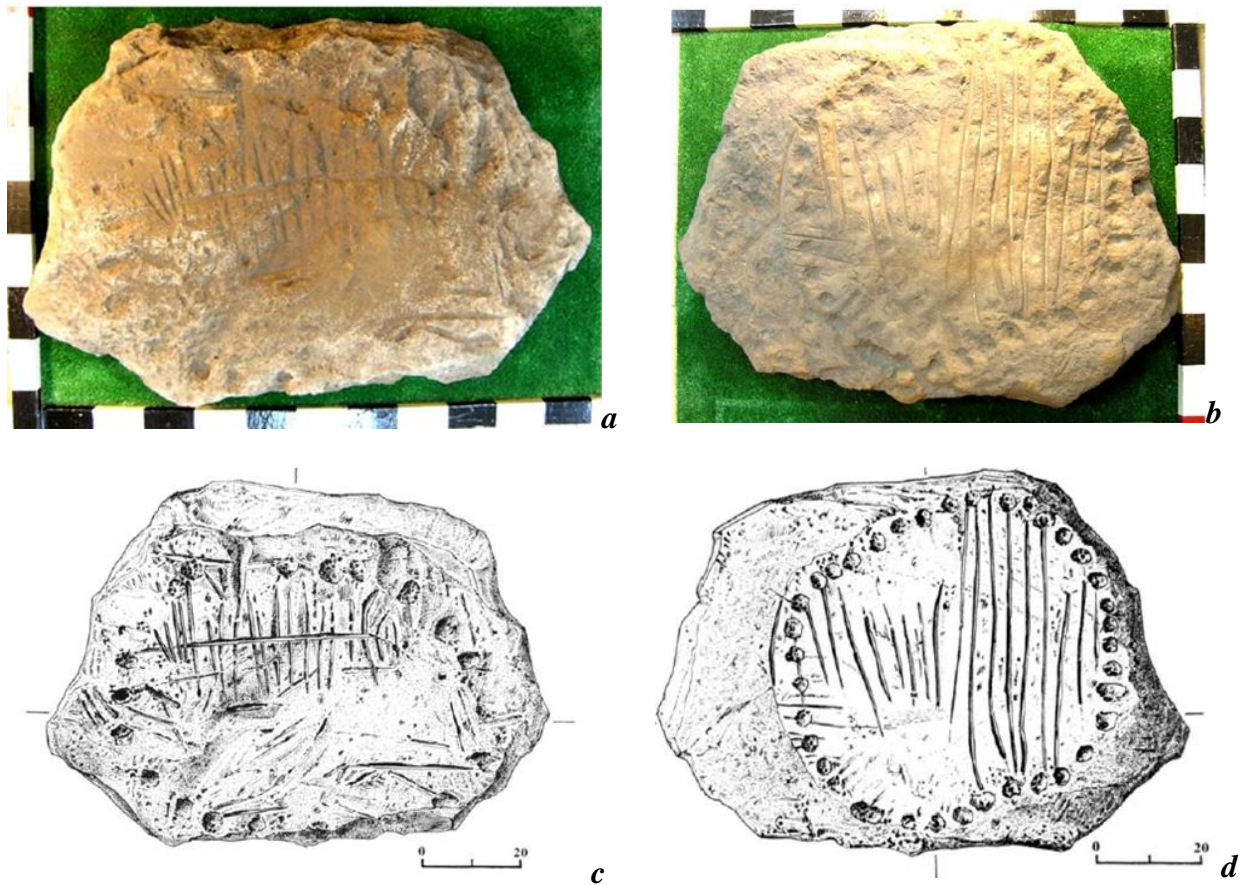

**Figure 6.** Popov Yar-2, with cup marks: *a* – photo of side A (Polidovich et al., 2012, fig. 214), *b* – photo of side B (Polidovich et al., 2012, fig. 218), *c* – drawing of side A (Polidovich et al., 2012, fig. 212), *d* – drawing of side B (Polidovich et al., 2012, fig. 212).

Another slab with cup marks located in a semicircle was discovered in the Krasnodar Krai near a heavily plowed mound near the settlement Pyatikhatki of Anapa District (Novichikhin, 1995).

It was attributed to the Dolmen culture and dated to the XXV-XV centuries BC. Now it is kept in the Anapa Archaeological Museum (Fig. 7).

Studies of the above slabs have shown that the cup marks on these slabs represent the hour markers of the "dial" of an analemmatic sundial (Vodolazhskaya, 2013; Vodolazhskaya, Larenok, Nevsky, 2014; Vodolazhskaya, Novichikhin, Nevsky, 2021).

The cup marks of the "Sun Stone" are arranged in a circle, as in the case of the slabs from the Northern Black Sea region. Therefore, it has been suggested that the slab from Montenegro may also be an analemmatic sundial.

In the case of analemmatic sundials, the cup marks must be located along the arc of the ellipse and must be in the northern part of the slab necessarily. The slab itself should be located horizontally, and the gnomon should be installed vertically and move according to the principle of a chess piece or suspended over the central part of the slab with the ability to move along the North-South line (Savoie, 2009, p. 111-123).

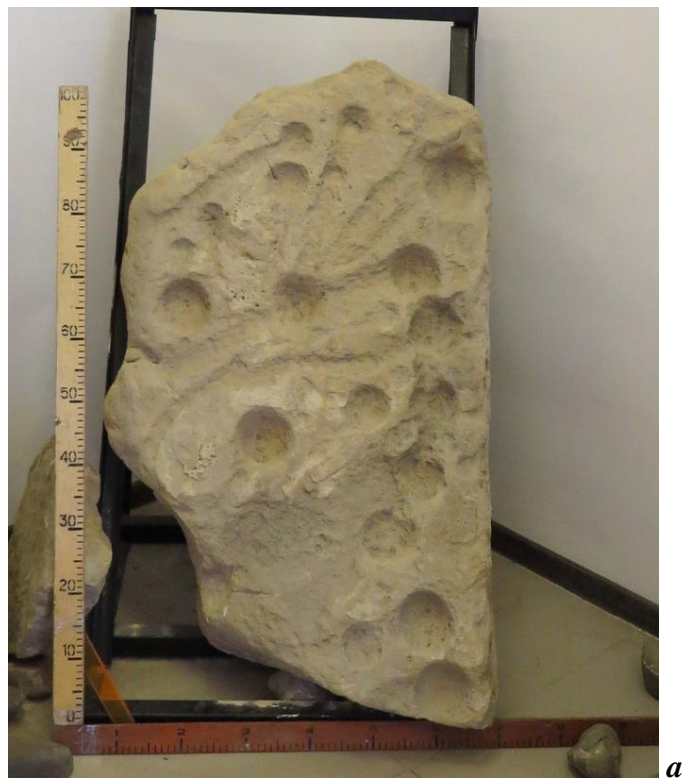

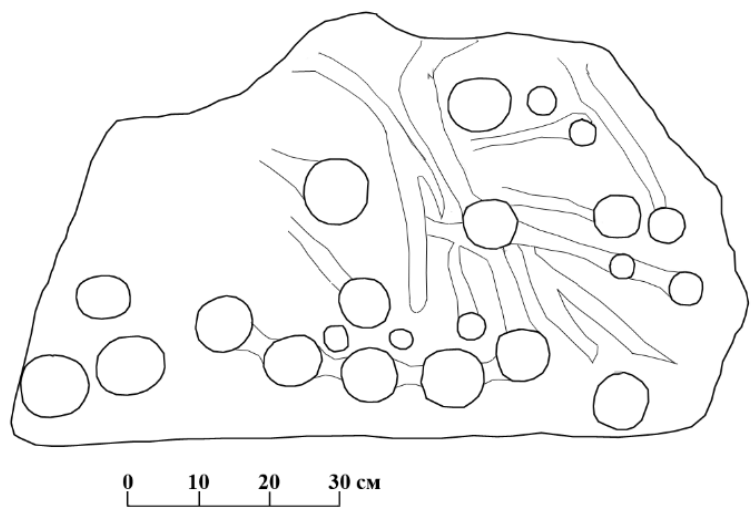

**Figure 7.** Slab from the settlement Pyatikhatki: ***a*** – photo in the modern exposition of the Anapa Archaeological Museum (Vodolazhskaya, Novichikhin; Nevsky, 2021, fig. 1); ***b*** – schematic drawing of the slab surface (Vodolazhskaya, Novichikhin; Nevsky, 2021, fig. 3).

The "Sun Stone" is more similar in markings to the Srubnaya culture slabs: all the cup marks are small and are located along the line of the ellipse, so we preliminarily dated it to the XV-XII centuries BC. The cultural identity of the "Sun Stone" remains unclear.

Based on the preliminary dating, the coordinates of hour markers of the analemmatic sundial were calculated for 1500 BC for the geographical coordinates of the place of detection 42°31′09″N, 19°46′13″E. For the calculation the same method was used as for the slabs from the Northern Black Sea region (see, for example, Vodolazhskaya, Novichikhin, Nevsky, 2021).

$$M = \frac{m}{\sin \varphi}, \quad (1)$$

$$x = M \cdot \sin H, \quad (2)$$

$$y = M \cdot \sin \varphi \cdot \cos H, \quad (3)$$

$$Z_{ws} = M \cdot tg\delta_{ws} \cdot \cos\varphi, \quad (4)$$

$$Z_{ss} = M \cdot tg\delta_{ss} \cdot \cos\varphi, \quad (5)$$

$$H' = arctg(tgH/\sin\varphi), \text{ при } t \in [6; 18] \quad (6)$$

$$H' = arctg(tgH/\sin\varphi) - 180°, \text{ при } t \in [0; 6[$$

$$H' = arctg(tgH/\sin\varphi) + 180°, \text{ при } t \in ]18; 24],$$

где $H = 15° \cdot (t - 12),$

where $x$ – the coordinate of a point along the $X$ axis for an analemmatic sundial, $y$ – the coordinate of a point along the $Y$ axis for an analemmatic sundial, $m$ – the semi-minor axis of the ellipse, $M$ – the semi-major axis of the ellipse, $\varphi$ – the latitude of the area, $t$ – the true local solar time, $H$ – the hour angle of the Sun, $H'$ – the angle between the noon line and the hour line on the clock relative to the center of coordinates (center of the ellipse), $\delta_{ws}=-\varepsilon$ – declination of the Sun on the day of the winter solstice, $\delta_{ss}=\varepsilon$ – declination of the Sun on the day of the summer solstice, $y=Z_{ws}$ – on the day of the winter solstice, $y=Z_{ss}$ – on the day of the summer solstice (Fig. 8). On the days of the equinox $\delta_{eq}=0$.

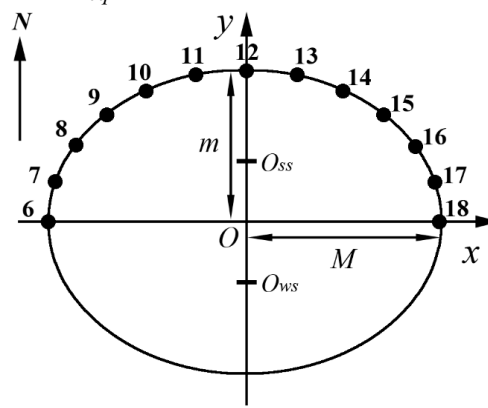

**Figure 8.** Coordinate plane with hour markers from 6 to 18 o'clock. M is the semi-major axis of the ellipse, *m* is the semi-minor axis of the ellipse, *O* is the center of the ellipse, $O_{ws}$ is the position of the gnomon on the winter solstice for analemmatic clock, $O_{ss}$ is the position of the gnomon on the summer solstice for the analemmatic clock.

Three grooves are clearly visible on the surface of the "Sun Stone". If we extend their linear fragments going from the located elliptically cup marks to the center, then they will intersect at some point O approximately in the center of the slab (Fig. 9). Assuming that the center of the ellipse was set in this way for marking hour markers, we built a horizontal coordinate system centered at point O and the OX axis passing along the central groove.

However, in this case, the OY axis did not pass through the cup mark, which should correspond to 12 o'clock, but somewhat to the right of it. Because the grooves defining the center of coordinates were not perfectly linear, it was concluded that the location of the center may not be very accurate. This conclusion allowed us to slightly shift the OY axis to the left – so that it passes through the center of the nearest cup mark – presumably 12 o'clock (Fig. 10).

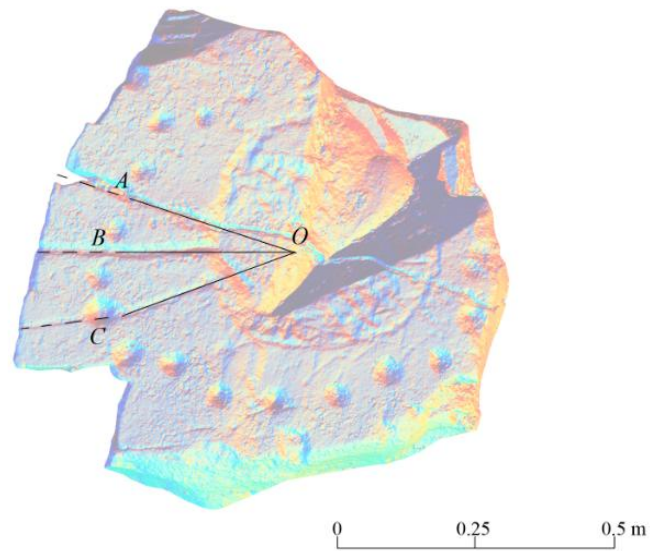

**Figure 9.** Reconstruction of the center of the slab marking ellipse using grooves.

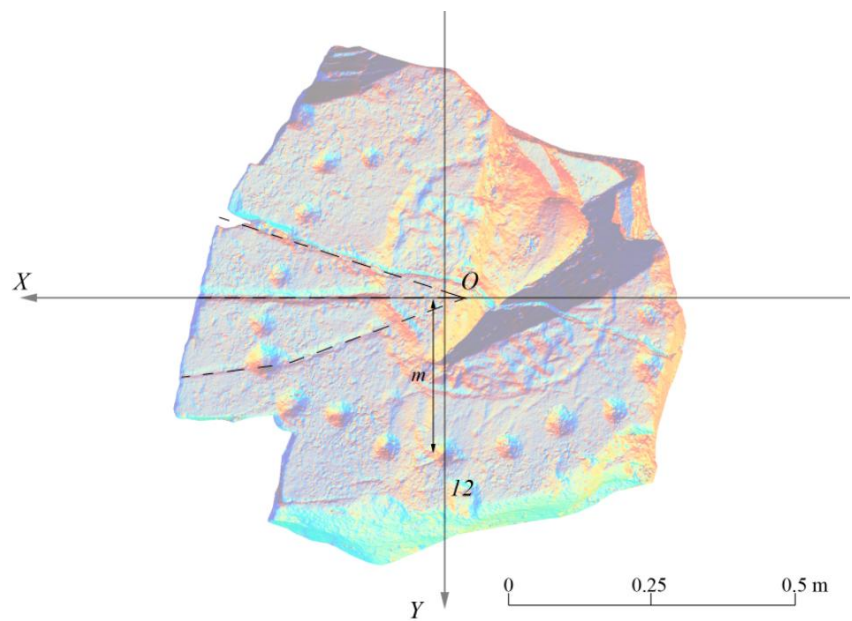

**Figure 10.** Reconstruction of the center of the slab marking ellipse, taking into account the central cup mark, presumably corresponding to 12 o'clock.

With this location of the center of coordinates, the measured semi-minor axis m of the ellipse of the hour markings turned out to be ≈27 cm relative to the center of the supposed cup mark at 12 o'clock.

Calculated for the minor semi axis m = 27 cm, the value of the major semiaxis M = 39.95≈40 cm, the displacement of the gnomon along the Y-axis on the day of the winter solstice $Z_{ws}$ = -13.05≈-13 (cm), and on the day of the summer solstice $Z_{ss}$≈13 (cm). Those to correctly measure the time, the gnomon had to be shifted on the days of the summer solstice by ≈13 cm north of the center of coordinates (point $O_{ws}$), and on days of the winter solstice by ≈13 cm to the south of the acnter of coordinates (point ($O_{ss}$) (Fig. 11). When constructing lines parallel to the OX axis and passing through the $O_{ws}$ and $O_{ss}$ points, it becomes clear that the grooves, that define the center of coordinates, with their outer edges mark the distances by which the gnomon should be shifted on the days of the summer and winter solstices.

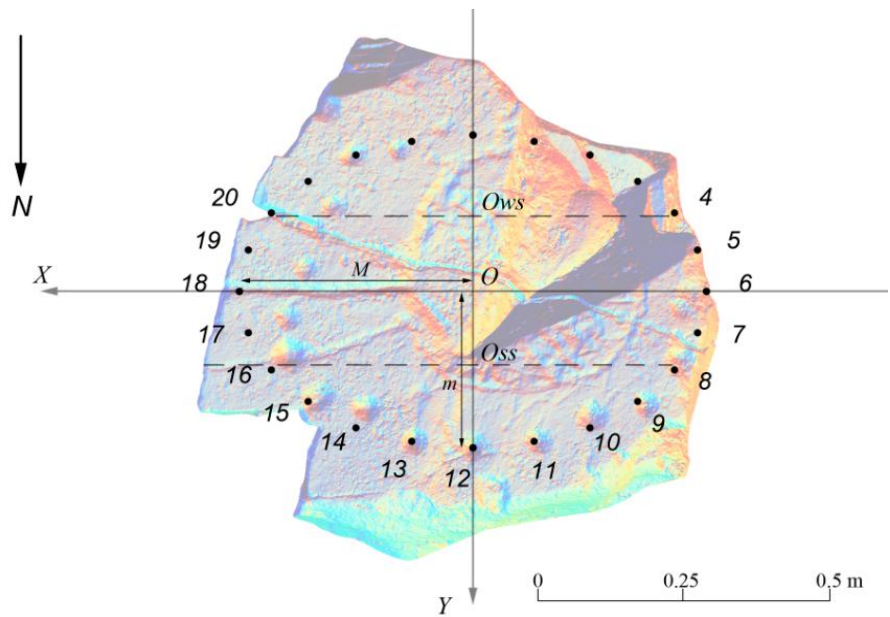

**Figure 11.** The projection of the calculated hour markers on the surface of the "Sun Stone" in accordance with the measured value of the minor semiaxis m and the reconstructed center of coordinates *O*. *N* is the direction to the true North in accordance with the calculated parameters of the hour marking ellipse.

The results of calculating the coordinates x and y of the hour marks of the analemmatic clock for the latitude of the "Sun Stone" detection are given in Table 1.

**Table 1.** Calculated coordinates of hour markers of the analemmatic sundial for latitude 44°58′ N: H – the hour angle of the Sun, H' – the angle between the noon line and the hour line on the sundial, t – the mean solar time, x – the coordinate of the hour markers along the X axis, y – the y-coordinate of the clock marks.

|  | *T, (hour)* | | | | | | | | | | | | |
|---|---|---|---|---|---|---|---|---|---|---|---|---|---|
|  | 0 | 1 | 2 | 3 | 4 | 5 | 6 | 7 | 8 | 9 | 10 | 11 | 12 |
| H, (°) | -180,0 | -165,0 | -150,0 | -135,0 | -120,0 | -105,0 | -90,0 | -75,0 | -60,0 | -45,0 | -30,0 | -15,0 | 0,0 |
| H', (°) | -180,0 | -159,2 | -140,8 | -125,2 | -112,2 | -100,7 | -90,0 | -79,3 | -67,8 | -54,8 | -39,2 | -20,8 | 0,0 |
| x, (cm) | 0,0 | -10,3 | -20,0 | -28,3 | -34,6 | -38,6 | -40,0 | -38,6 | -34,6 | -28,3 | -20,0 | -10,3 | 0,0 |
| y, (cm) | -27,0 | -26,1 | -23,4 | -19,1 | -13,5 | -7,0 | 0,0 | 7,5 | 14,5 | 20,5 | 25,1 | 28,0 | 29,0 |

|  | *t, (hour)* | | | | | | | | | | | | |
|---|---|---|---|---|---|---|---|---|---|---|---|---|---|
|  | 12 | 13 | 14 | 15 | 16 | 17 | 18 | 19 | 20 | 21 | 22 | 23 | 24 |
| H, (°) | 0,0 | 15,0 | 30,0 | 45,0 | 60,0 | 75,0 | 90,0 | 105,0 | 120,0 | 135,0 | 150,0 | 165,0 | 180,0 |
| H', (°) | 0,0 | 20,8 | 39,2 | 54,8 | 67,8 | 79,3 | 90,0 | 100,7 | 112,2 | 125,2 | 140,8 | 159,2 | 180,0 |
| x, (cm) | 0,0 | 10,3 | 20,0 | 28,3 | 34,6 | 38,6 | 40,0 | 38,6 | 34,6 | 28,3 | 20,0 | 10,3 | 0,0 |
| y, (cm) | 27,0 | 26,1 | 23,4 | 19,1 | 13,5 | 7,0 | 0,0 | -7,5 | -14,5 | -20,5 | -25,1 | -28,0 | -29,0 |

When applying the calculated hour markers to the plan-diagram of the surface of the "Sun Stone", it can be seen that they coincide well with the cup marks on the slab, with the exception of the easternmost ones, corresponding to the range from 17:00 to 21:00. True, in this case, the hour markers are still close to the corresponding cup marks, and the function of the 18 o'clock cup mark is taken over by the central groove.

To correctly measure the time on the days of the equinox, the gnomon had to be set at point O (Fig. 12). The day began at 6 o'clock and ended at 18 o'clock. The corresponding hour lines are marked in green in the figure.

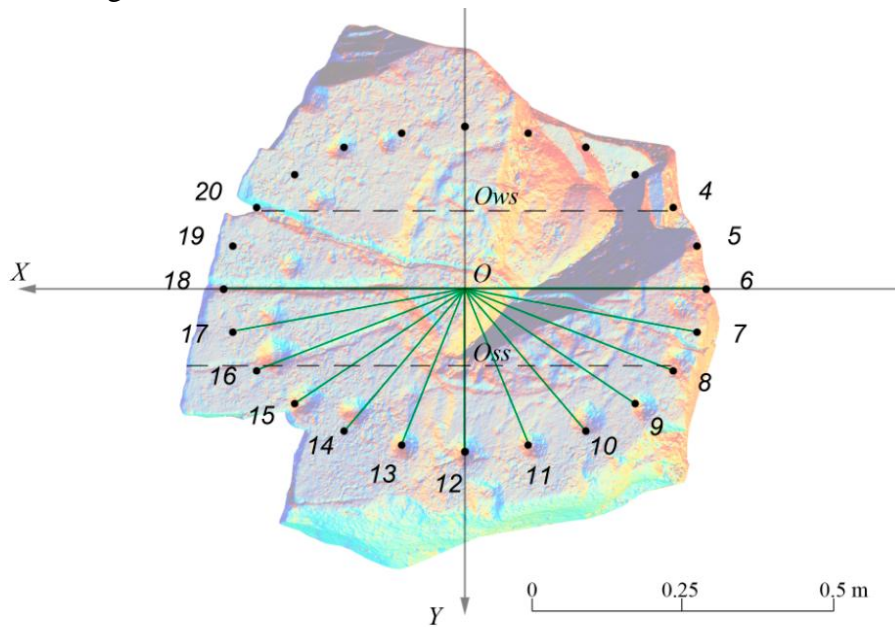

**Figure 12.** Hour lines corresponding to the position of the gnomon at the equinox.

To correctly measure the time on the day of the winter solstice, the gnomon had to be set at the Ows point (Fig. 13). The day began at about 7:30 am and ended at 4:30 pm (the times of sunrise and sunset were calculated using the RedShift 7 program). The corresponding hour lines are marked in blue in the figure.

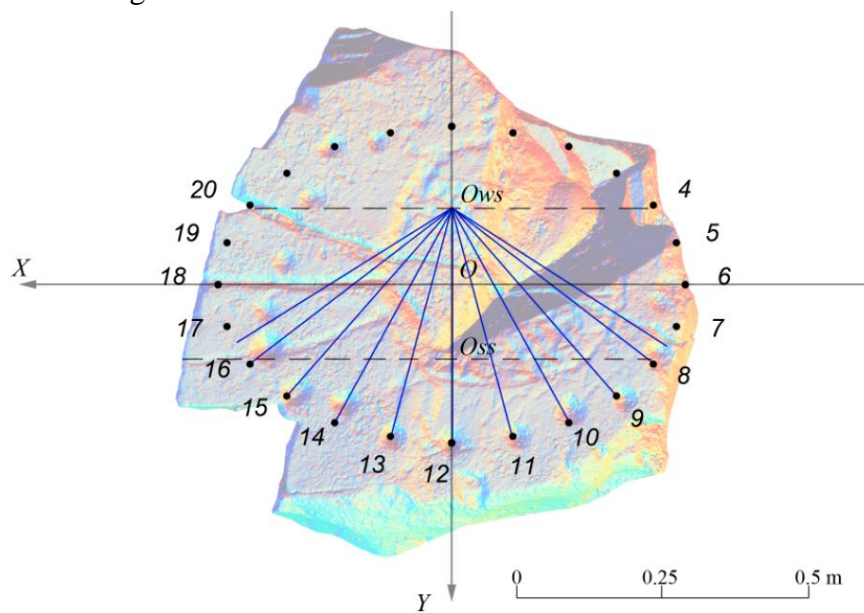

**Figure 13.** Hour lines corresponding to the position of the gnomon at the winter solstice.

To correctly measure the time on the day of the summer solstice, the gnomon had to be installed at the Oss point (Fig. 14). The day began at about 4:23 pm and ended at 7:37 pm (the times of sunrise and sunset were calculated using the program RedShift 7 Advanced). The corresponding hour lines are marked in red in the figure.

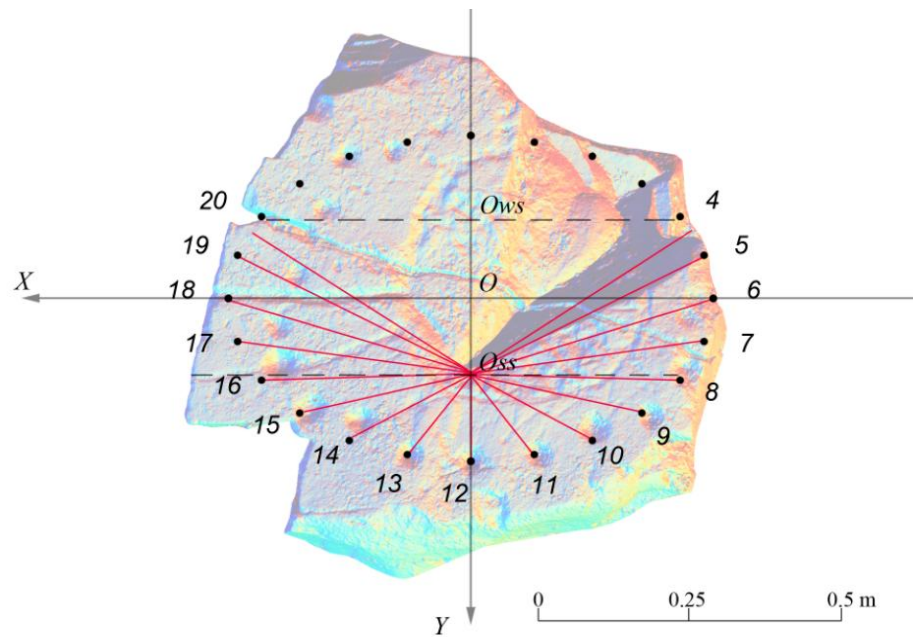

**Figure 14.** Hour lines corresponding to the position of the gnomon at the summer solstice.

With such an arrangement of hour markers on the surface of the "Sun Stone", it becomes clear why a central cleavage could have formed. It is most likely that there was a groove on the surface of the slab that coincided with the hour line for the time of sunrise on the day of the summer solstice. They could try to expand it for technological or ritual reasons, or for this groove (along it) a deliberate blow was made for ritual reasons. This led to the formation of a chip, the trace from which also testifies in favor of the interpretation of the "Sun Stone" as an analemmatic sundial. We would also like to note that the ellipses of two of the three analemmatic sundials from the Northern Black Sea region are close in size to the reconstructed ellipse of the "Sun Stone" (table 2).

**Table 2.** Parameters of the ellipse marking the analemmatic sundials: m – semi-minor axis of the ellipse, M – semi-major axis of the ellipse, Z – the distance by which the gnomon moves on the days of the solstices.

| Slab detection location | m, cm | M, cm | Z, cm |
|---|---|---|---|
| Maja e Can, Prokletije (Western Balkans) | 27,0 | 39,9 | 13,1 |
| Kurgan group Popov Yar-2 (Northern Black Sea region) | 28,4 | 38,0 | 11,1 |
| Settlement Pyatikhatki (Northern Black Sea region) | 29,0 | 41,0 | 12,9 |
| Burial ground Tavria-1 (Northern Black Sea region) | 14,0 (x2=28,0) | 19,0 (x2=38,0) | 5,7 (x2=11,4) |

On the Popov Yar-2 slab, the groove marks the distance the gnomon should move on the day of the winter solstice, similar to the groove on the "Sun Stone".

Thus, the marking of the analemmatic sundial on the slab from Prokletiy (Montenegro) is similar in its parameters and features to the marking of the analemmatic sundial on the Srubnaya culture slabs from the Northern Black Sea region. This confirms the preliminary dating of the slab of the XV-XII centuries BC and testifies to the contacts of representatives of the local

Glasinac culture (Glasinac II - Glasinac III) of the Western Balkans with representatives of the Srubnaya culture of the Northern Black Sea region.

Acknowledgments

The authors would like to thank Ida Ferdinandi for her help in making this research possible and Ilijaz Duraković from Gusinje for providing valuable information which helped the research and discovery of the sundial.